\RequirePackage{amsmath}
\documentclass[letterpaper, 10 pt, conference]{ieeeconf}  % Comment this line out
\IEEEoverridecommandlockouts                              % This command is only
\usepackage{algorithm}
\usepackage{algpseudocode}
\usepackage{setspace}
\usepackage{graphicx}
\usepackage[caption=false]{subfig}
\usepackage{multirow}
\usepackage{url}
\usepackage{mathtools}
\usepackage{amssymb}
\usepackage{float}
\usepackage{hhline}

\DeclarePairedDelimiterX\Basics[1](){ #1}

\title{\LARGE \bf
Bit-Vector Model Counting using Statistical Estimation
}

\author{%{\em Working paper as of August 29th, 2017}\\
    Seonmo Kim and Stephen McCamant
	\thanks {}\\
    Department of Computer Science \\ 
    University of Minnesota\\
    Minneapolis, MN 55455, USA\\
}% <-this % stops a space

\begin{document}

\maketitle
\thispagestyle{plain}
\pagestyle{plain}

%%%%%%%%%%%%%%%%%%%%%%%%%%%%%%%%%%%%%%%%%%%%%%%%%%%%%%%%%%%%%%%%%%%%%%%%%%%%%%%%
\begin{abstract}

Approximate model counting for bit-vector SMT formulas (generalizing \#SAT)
has many applications such as probabilistic inference and quantitative
information-flow security, but it is computationally difficult.
Adding random parity constraints (XOR streamlining) and then checking
satisfiability is an effective approximation technique, but it requires
a prior hypothesis about the model count to produce useful results.
We propose an approach inspired by statistical estimation to continually
refine a probabilistic estimate of the model count for a formula, so that
each XOR-streamlined query yields as much information as possible.
We implement this approach, with an approximate probability model, as
a wrapper around an off-the-shelf SMT solver or SAT solver.
Experimental results show that the implementation is faster than the most
similar previous approaches which used simpler refinement strategies.
The technique also lets us model count formulas over floating-point
constraints, which we demonstrate with an application to a vulnerability
in differential privacy mechanisms.

\end{abstract}

\begin{keywords}
model counting, bit-vectors, floating point, \#SAT, randomized algorithms
\end{keywords}

%%%%%%%%%%%%%%%%%%%%%%%%%%%%%%%%%%%%%%%%%%%%%%%%%%%%%%%%%%%%%%%%%%%%%%%%%%%%%%%%
\section{Introduction}\label{sec:intro}

Model counting is the task of determining the number of satisfying
assignments of a given formula.
Model counting for Boolean formulas, \#SAT, is a standard
model-counting problem, and it is a complete problem for the complexity
class \#P in the same way that SAT is complete for NP.
\#P is believed to be a much harder complexity class than NP, and
exact \#SAT solving is also practically much less scalable than SAT
solving.
\#SAT solving can be implemented as a generalization of the DPLL
algorithm~\cite{Davis:1962:MPT:368273.368557}. and a number of systems
such as \textsf{{Relsat}}~\cite{Bayardo:1997:UCL:1867406.1867438},
\textsf{{CDP}}~\cite{DBLP:journals/jair/BirnbaumL99},
\textsf{{Cachet}}~\cite{DBLP:conf/sat/SangBBKP04},
\textsf{{sharpSAT}}~\cite{Thurley_sharpsat},
\textsf{{DSHARP}}~\cite{Muise_dsharp} and \textsf{{countAntom}}~\cite{Burchard2015} have demonstrated various
optimization techniques.
However, not surprisingly given the problem's theoretical hardness,
such systems often perform poorly when formulas are large and/or have
complex constraints.

Since many applications do not depend on the model count being exact,
it is natural to consider approximation algorithms that can give an
estimate of a model count with a probabilistic range and confidence.
Some approximate model counters include
\textsf{{ApproxCount}}~\cite{Wei:2005:NAM:2129929.2129954},
\textsf{{SampleCount}}~\cite{Gomes07fromsampling},
\textsf{{MiniCount}}~\cite{Kroc_leveragingbelief},
\textsf{{ApproxMC}}~\cite{DBLP:conf/cp/ChakrabortyMV13},
\textsf{{ApproxMC-p}}~\cite{QAPL2016} and
\textsf{{ApproxMC2}}~\cite{CMV16}
.
In this paper we build on the approximation technique of XOR
streamlining~\cite{DBLP:conf/aaai/GomesSS06}, which reduces the number
of solutions of a formula by adding randomly-chosen XOR (parity)
constraints.
In expectation, adding one constraint reduces the model count by a
factor of 2, and $k$ independent constraints reduce the model count by
$2^k$.
If a formula with extra constraints has $n > 0$ solutions, the
original formula likely had about $n\cdot 2^k$.
If the model count after constraints is small, it can be found with a
few satisfiability queries, so XOR streamlining reduces approximate
model counting to satisfiability.
However to have an automated system, we need an approach to choose
$k$ values when the model count is not known in advance.
 
One application of approximate model counting is measuring the amount
of information revealed by computer programs.
For a deterministic computation, we say that the {\em
  influence}~\cite{Newsome:2009:MCC:1554339.1554349} is the base-two
log of the number of distinct outputs that can be produced by varying
the inputs, a measure of the information flow from inputs to outputs.
Influence computation is related to model counting, but formulas
arising from software are more naturally expressed as SMT
(satisfiability modulo theories) formulas over bit-vectors than as
plain CNF, and one wants to count values only of output variables
instead of all variables.
The theory of arithmetic and other common operations on bounded-size
bit-vectors has the same theoretical expressiveness as SAT, since
richer operations can be expanded (``bit-blasted'') into circuits.
But bit-vector SMT is much more convenient for expressing the computations
performed by software, and SMT solvers incorporate
additional optimizations.
We build a system for this generalized version of the problem which
takes as input an SMT formula with one bit-vector variable
designated as the output, and a specification of the desired
precision.
%
% By repeated SMT queries using XOR streamlining, our system produces a
% range estimate consisting of a lower and an upper bound that are
% close to each other and usually bound the true influence.

Our algorithm takes a statistical estimation approach.
It maintains a probability distribution that reflects an estimate of
possible influence values, using a particle filter consisting of
weighted samples from the distribution.
Intuitively the mean of the distribution corresponds to our
tool's best estimate, while the standard deviation becomes
smaller as its confidence grows.
At each step, we refine this estimate by adding $k$ XOR constraints to
the input formula, and then enumerating solutions under those
constraints, up to a maximum of $c$ solutions (we call this
enumeration process an {\em exhaust-up-to-$c$}
query~\cite{Newsome:2009:MCC:1554339.1554349}).
At a particular step, we choose $k$ and $c$ based on our previous
estimate (prior), and then use the query result to update the estimate
for the next step (posterior).
The update from the query reweights the particle filter points
according to a probability model of how many values are excluded
by XOR constraints.
We use a simple binomial-distribution model which would be exact if each
XOR constraint were fully independent.
Because this model is not exact, a technique based only on it does not
provide probabilistic soundness, even though it performs well practically.
So we also give a variant of our technique which does produce a sound bound,
at the expense of requiring more queries to meet a given precision goal.

We implement our algorithm in a tool \textsf{SearchMC}\footnotemark\ that wraps
either a
bit-vector SMT solver compatible with the SMT-LIB 2 standard or a
SAT solver, and report experimental results.
\footnotetext{The source code is available at \url{https://github.com/seonmokim/SearchMC}}
\textsf{SearchMC} can be used to count solutions with respect to a subset of the
variables in a formula, such as the outputs of a computation, the
capability that Klebanov et al. call projected model counting~\cite{QAPL2016},
and Val et al. call subset model counting~\cite{ValEBAH2016}.
%
%This capability is also used by Irvii et al.~\cite{IvriiMMV2016} to
%accelerate \#SAT by not counting over redundant variables.
%
In our case the variables not counted need not be of bit-vector type.
For instance this makes \textsf{SearchMC} to our knowledge the first tool that can
be used to count models of constraints over floating-point numbers
(counting the floating-point bit patterns individually, as contrasted with
computing the measure of a subset of $\mathbb{R}^n$ as in the work of
Chistikov et al.~\cite{ChistikovDM2015}).
We demonstrate the use of this capability with an application to a
security problem that arises in differential privacy mechanisms because
of the limited precision of floating-point values.

Compared to \textsf{{ApproxMC2}}~\cite{CMV16} and
\textsf{{ApproxMC-p}}~\cite{QAPL2016},
concurrently-developed approximate \#SAT tools also based on XOR streamlining,
our technique
gives results more quickly for the same requested confidence levels.
%
% Our tool's primary probabilistic results are empirical, not provably sound
% as \textsf{{ApproxMC2}}'s, but we describe how our tool can separately compute
% sound though less precise bounds from the same query results.
%
%We also compare to a state-of-the-art exact \#SAT solver,
%\textsf{{countAntom}}~\cite{Burchard2015}, and show better performance
%on some classes of problem as size increases.
%
%We also demonstrate the application to quantitative information
%flow by measuring the influence of some image anonymization
%transformations.

In summary, the key attributes of our approach are as follows:
\begin{itemize}
    \item Our approximate counting approach gives a two-sided bound with user-specified confidence.
    \item Our tool inherits the expressiveness and optimizations of SMT solvers.
    \item Our tool gives a probabilistically sound estimate if requested, or can
    give a result more quickly if empirical precision is sufficient.
	%\item Given a prior distribution and a SAT/SMT query result, our tool computes a posterior distribution based on simulated experiments.
	%\item Our tool's probability estimate gives bounds even if interrupted.
\end{itemize}

\section{Background}\label{sec:background}
\textbf{\textit{XOR Streamlining.}} The main 
idea of XOR streamlining~\cite{DBLP:conf/aaai/GomesSS06} is to add randomly chosen XOR constraints to a given 
input formula and feed the augmented formula to a satisfiability solver. One random XOR 
constraint will reduce the expected number of solutions in half. Consequently, if the formula is still satisfiable
after the addition of $s$ XOR constraints, the original formula likely has at least $2^s$ models. 
If not, the formula likely has at most $2^s$ models. Thus we can obtain a lower bound or an upper bound
with this approach.
There are some crucial parameters to determine the bounds 
and the probability of the bounds and they need to be carefully chosen in order to obtain good bounds.
However, early systems~\cite{DBLP:conf/aaai/GomesSS06} did not provide an algorithm to choose the parameters.\\
\textbf{\textit{Influence.}} Newsome \textit{et al.}~\cite{Newsome:2009:MCC:1554339.1554349} introduced the terminology
of ``influence'' for a specific application of model counting in quantitative information-flow measurement. This idea can capture the control of input variables over an output
variable and distinguish true attacks and false positives in a scenario of malicious input to a network service. 
The influence of input variables over
an output variable is the $\log_2$ of the number of possible output values.\\
\textbf{\textit{Exhaust-up-to-$c$ query.}} Newsome \textit{et al.}\@ also introduced the terminology of an ``exhaust-up-to-$c$ query'',
which repeats a satisfiability query up to some number $c$ of solutions, or until there are no satisfying values left. This is
a good approach to find a model count if the number of solution is small.\\ 
\textbf{\textit{Particle Filter.}}
A particle filter~\cite{CIS-135379} is an approach to the statistical estimation of
a hidden state from noisy observations, in which a probability distribution over the
state is represented non-parametrically by a collection of weighted samples referred to
as particles.
The weights evolve over time according to observations; they tend to become unbalanced,
which is corrected by a resampling process which selects new particles with balanced
weights.
A particle filtering algorithm with periodic resampling takes the following form:
\begin{itemize}
    \item [1.] Sample a number of particles from a prior distribution.
    \item [2.] Evaluate the importance weights for each particle and normalize the weights.
	\item [3.] Resample particles (with replacement) according to the weights.
	\item [4.] The posterior distribution represented by the resampled particles becomes the prior distribution to next round and go to step 2.
\end{itemize}

\section{Design}\label{sec:design}

This section describes the approach and algorithms used by \textsf{{SearchMC}}.
It is implemented as a wrapper around an off-the-self bit-vector
satisfiability solver that supports the SMT-LIB2 
format~\cite{Barrett10c.:the}.
It takes as input an SMT-LIB2 formula in a quantifier-free theory that includes
bit-vectors
(QF\_BV, or an extension like QF\_AUFBV or QF\_FPBV) 
in which one bit-vector is designated as the output, i.e. the bits over which
solutions should be counted.
(For ease of comparison with \#SAT solvers, \textsf{{SearchMC}} also has
a mode that takes a Boolean formula in CNF, with a list of CNF variables
designated as the output.)
\textsf{{SearchMC}} repeatedly queries the SMT solver with variations of the
supplied input which add XOR constraints and/or ``blocking'' constraints that
exclude previously-found solutions; based
on the results of these queries, it estimates the total number of values of
the output bit-vector for which the formula has a satisfying assignment.
%
%In the experiments reported in this paper, we used Z3~\cite{DeMoura:2008:ZES:1792734.1792766} as the SMT
%solver unless another solver is mentioned.

\textsf{{SearchMC}} chooses fruitful queries by keeping a running estimate of
possible values of the model count.
We model the influence ($\log_2$ of model count) as if it were a continuous
quantity, and represent the estimate as a probability distribution over possible
influence values.
In each iteration we use the current estimate to choose a query, and then
update the estimate based on the query's results.
(At a given update, the most recent previous distribution is called the
\textit{prior}, and the new updated one is called the \textit{posterior}.)
As the algorithm runs, the confidence in the estimate goes up, and the best
estimate changes less from query to query as it converges on the correct result.
Each counting query \textsf{{SearchMC}} makes is parameterized by $k$, the
number of random XOR constraints to add, and $c$, the maximum number of solutions
to count.
The result of the query is a number of satisfying assignments between 0 and $c$
inclusive, where a result that stops at $c$ means the real total is at least $c$.
Generally a low result leads to the next estimate being lower than the
current one and a high result leads to the estimate increasing.
Section~\ref{sec:updating} describes the process of updating the probability 
distribution, and then Section~\ref{sec:algorithm} gives the details of the
algorithms that use it.

\subsection{Updating distribution and confidence interval}\label{sec:updating}

We here explain the idea of how we compute a posterior distribution over influence, 
where both the prior and posterior are represented by particles.
Suppose we have a formula $f$ with a known influence $\log_2N$, and add $k$ XOR random constraints to the formula. If we simulate checking the satisfiability of this augmented formula $f'$ for different XOR constraints, we can estimate a probability of sat/unsat on $f'$.
We expand this idea by applying exhaust-up-to-$c$ approach to $f'$. We count the number of satisfying assignments $n$ up to $c$ and generate the distributions for each number of satisfying assignments (where $n=c$ means that the number
of satisfying assignments is in fact $c$ or more).
Thus under an assumption on the true influence of a formula, we can estimate the probabilities of 
each number of satisfying assignments based on $k$. By collecting these probabilities across a range of influence, we obtain 
a probability distribution over influence for an unknown formula assumed to have less than a maximum bits of influence.
Under the idealized assumption that each XOR constraint is completely
independent, adding $k$ XOR constraints will leave
each satisfying assignment alive with probability $1/2^k$. 
For any particular set of $n\ge 0$ satisfying assignments remaining out of an
original $N$, the probability that exactly those $n$ solutions will remain
is the product of $1/2^k$ for each $n$ and $1-(1/2^k)$ for each of the
other $N-n$.
Summing the total number of such sets with a binomial coefficient, we can
approximately model the probability of exactly $n$ solutions remaining as:

\begin{equation}
	Pr_{=n}(N,k)=\binom{N}{n}(\frac{1}{2^k})^{n}(1-\frac{1}{2^k})^{N-n} 
\end{equation}

For the case when the algorithm stops looking when there might still
be more solutions, we also want an expression for the probability
that the number of solutions is $n$ or more.
We compute this straightforwardly as one minus the sum of the probabilities
for smaller values:

\begin{equation}
	Pr_{\ge n}(N,k)=1-\sum_{i=0}^{n-1}Pr_{=i}(N,k)
\end{equation}

We use XOR constraints that contain each counted bit with probability
one half, and are negated with probability one half.
(This is the same family of constraints used in other recent
systems~\cite{DBLP:conf/cp/ChakrabortyMV13,QAPL2016,CMV16}.
Earlier work~\cite{DBLP:conf/aaai/GomesSS06} suggested using constraints
over exactly half of the bits, which have the same expected size, but
less desirable independence properties.)
Our binomial probability model is not precise in general, because these
XOR constraints are 3-independent, but not $r$-independent for $r\geq 4$.
When $N\geq 4$, some patterns among solutions (such as a set of four
bitvectors whose XOR is all zeros) lead to correlations in the
satisfiability of XOR constraints, and in turn to higher variance in
the probability distribution without changing the expectation.
This effect is relatively small, but we lack an analytic model of it,
so we compensate by slightly increasing the confidence level our
tool targets compared to what the user originally requested.

This probability model lets us simulate the probability of various query
results as a function of the unknown formula influence. We use this model as 
a weighting function for each particle and 
resample particles based on each particle's weight value. Then, we estimate
a posterior distribution from sampled particles that have all equal weights.
For instance, given a prior distribution over the influence sampled at
0.1 bit intervals, we can compute a sampled posterior distribution by
counting and re-normalizing just the probability weights that correspond
to a given query result value $n$.
From the estimated posterior distribution, the mean $\mu$ and the standard deviation $\sigma$ are computed. 
Hence, the $\mu$ is our best possible answer as our algorithm iterates and $\sigma$ shows how much
we are close to the true answer. 
Sequentially, the posterior distribution will be the next round's prior distribution and for use in the very first step of the algorithm we also implement
a case of the prior distribution as uniform over influence.

Next we compute a confidence interval (lower bound and upper bound) 
symmetrically from the mean of the posterior distribution 
even though the distribution is not likely to be symmetrical. 
There are several ways to compute the confidence interval but
the difference of the results is negligible as the posterior distribution gets narrower.
Therefore, we used a simple way to compute the confidence interval: a half interval from the left side of the mean and another half from the right side.

\subsection{Algorithm}\label{sec:algorithm}
We present our main algorithm \texttt{SearchMC} that runs automatically and always gives 
an answer with a given confidence interval. The pseudocode for algorithm \texttt{SearchMC} is 
given as Algorithm \ref{SearchMC}. 
Our algorithm takes as input a formula $f$, a desired confidence level $CL$ ($0<CL<1$), 
a confidence level adjustment $\alpha$ ($0\leq\alpha<1$), a desired range size $thres$ and an initial prior distribution $InitDist$.
$f$ contains a set of bit-vector variables and bit-vector operators. We can obtain a confidence interval at a confidence
level for a given mean and standard deviation. A confidence level $CL$ is a fraction parameter specifying the probability with which the interval should
contain the true answer, for example, 0.95 (95$\%$) or 0.99 (99$\%$). 
As we described above, the soundness of our model has not been proved hence a confidence level sometimes needs to be adjusted by the adjustment value $\alpha$ to meet the actual confidence level requirement.
If $\alpha=0$, we do not adjust the input confidence level. 
In our experiments, we used $\alpha=0.5$, which empirically leads
\textsf{{SearchMC}} to select large enough ranges. 
Finding the most appropriate value for $\alpha$ is future work.
Our algorithm terminates 
when the length of our confidence interval is less than or equal to a given non-negative parameter $thres$. This parameter 
determines the amount of running time and there is a trade-off. If $thres$ value is small, it gives a narrow confidence 
interval, but the running time would be longer. If the value is large, it gives a wide confidence interval, but a shorter running time.
Our tool can choose any initial prior distribution $InitDist$ represented by particles. For example, if we do not have any knowledge of input formula,
it might be better to start with a uniform distribution over 0 to a number of output variables. If we have a small knowledge that the true influence is less than 64, a uniform distribution over 0 to 64 might perform better.

\begin{algorithm}[!ht]
\caption{\texttt{SearchMC}($f$, $thres$, $CL$, $\alpha$, $InitDist$)}\label{SearchMC}
\begin{algorithmic}[1]
	%\State $c_{max} \gets 15$
    %\Comment The maximum $c$ value from our tables
    \State $CL \gets CL+(1-CL)\times \alpha$
    \Statex{\Comment Confidence level adjustment}
    %\State $max\_w\gets 64$
    %\Statex{\Comment The maximum width of an output bit-vector}
    \State $width\gets \texttt{getWidth}(f)$
	\Statex{\Comment The width of the output bit-vector of $f$}
    \State $prior \gets InitDist$
    \Comment Initial distribution
    \State $\delta \gets width$
    \While{$\delta>thres$}
    	\State $c,k\gets$ \texttt{ComputeCandK}$(prior, width)$
        \State $nSat\gets$ \texttt{MBoundExhaustUpToC}$(f, width, k, c)$
        \State $post, UB, LB\gets$ \texttt{Update}$(prior, c, k, nSat, CL)$
        \Statex{\Comment See Sec.~\ref{sec:updating}}
        \State $\delta \gets UB - LB$
        \If {$k == 0$}
        	\State output ``Exact Count: '', $nSat$
        \Else
        	\State $prior\gets post$
            \State output ``Lower: ", $LB$, ``Upper: '', $UB$
        \EndIf
    \EndWhile
\end{algorithmic}
\end{algorithm}

\begin{figure}[thpb]
      \centering
      \includegraphics[width=0.5\textwidth]{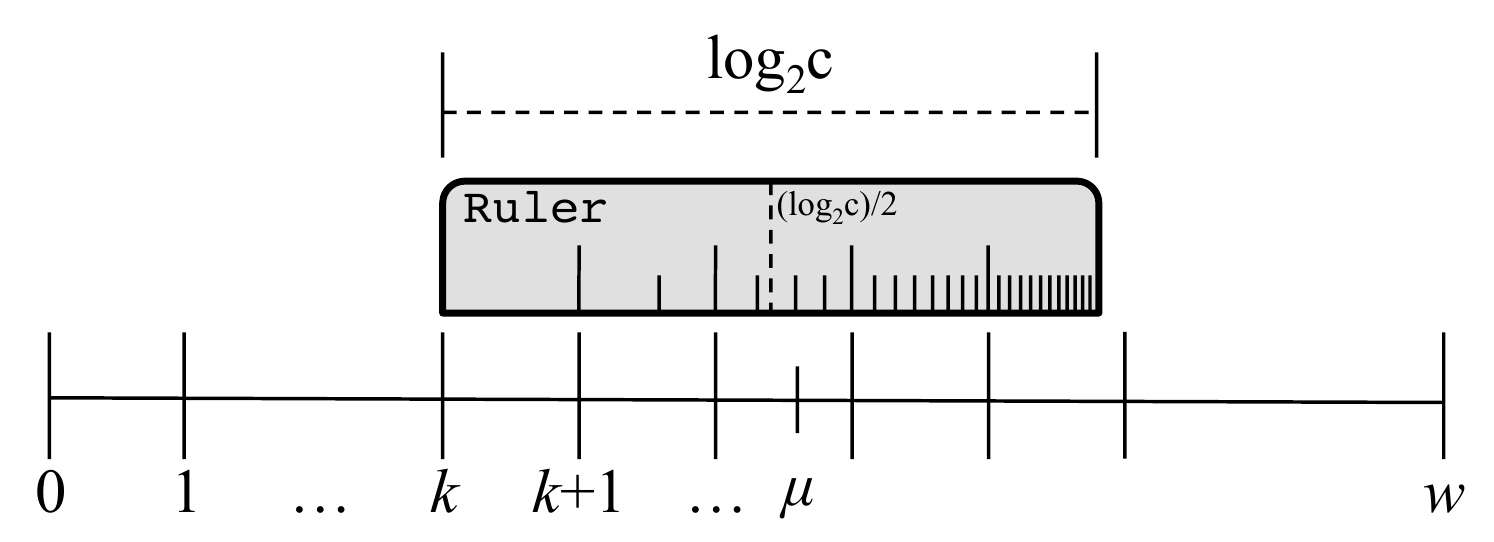}
      \caption{Ruler Intuition}
      \label{fig:Ruler}  
   \end{figure}

\begin{algorithm} [!ht]
\caption{\texttt{ComputeCandK}($prior$, $width$)}\label{ComputeCandK}
\begin{algorithmic}[1]
   \State $\mu, \sigma\gets$\texttt{getMuSigma}$(prior)$
   \State $c \gets \lceil((2^\sigma+1)/(2^\sigma-1))^2\rceil$
   %\If {$c > c_{max}$} 
   %\State $c \gets c_{max}$
   %\EndIf
   \State $k \gets \lfloor\mu-\frac{1}{2}\log_2c\rfloor$
   \If {$k \leq 0$}
   \State $c \gets 2^{width}+1$
   \Statex{\Comment In this case, $c$ is effectively infinite}
   \State $k \gets 0$
   \Comment No constraints
   \EndIf
   \State \Return $c,k$
\end{algorithmic}
\end{algorithm}
\noindent
\textbf{\textit{Variables.}} There are several variables: $prior$, $post$, $width$, $k$, $c$, 
$nSat$,  $UB$, $LB$ and $\delta$. $prior$ represents a prior distribution by sampled particles with corresponding weights. 
In one iteration, we obtain the updated posterior distribution $post$ with resampled 
particles based on our probabilistic model as described in Section~\ref{sec:updating}. 
The posterior becomes the prior distribution for the next iteration. 
While our algorithm is in the loop, it keeps updating $post$. 
%We start our initial prior distribution, sampled uniformly, to be a uniform distribution. 
$width$ is the width of the output bit-vector of an input formula $f$, which is an initial upper bound for the influence since the influence
cannot be more than the width of the output bit-vector.
$k$ is a number of random XOR constraints and $c$ specifies the maximum
number of solutions for the 
exhaust-up-to-$c$ query. 
%$c_{max}$ is a maximum value of $c$ such that $c \leq c_{max}$. $c_{max}$ is determined by the size of
%our empirical tables.
We obtain $c$ and $k$ using the \texttt{ComputeCandK} function shown as Algorithm~\ref{ComputeCandK}
and discussed below.
$nSat$ is a number of solutions from the exhaust-up-to-$c$ query. \texttt{MBoundExhaustUpToC} 
runs until it finds 
the model count exactly or $c$ solutions from formula $f$ with $k$ random XOR constraints.
$UB$ and $LB$ are variables to store
an upper bound and a lower bound of the current model count approximation with a given confidence level as we describe in Section~\ref{sec:updating}.
$\delta$ is the distance between
the upper bound and lower bound. This parameter determines whether our algorithm terminates or not. If $\delta$ is less than
or equal to our input value $thres$, our algorithm terminates. If not, it runs again with updated $post$ until $\delta$ reaches the desired range size $thres$. An extreme case $k=0$ denotes that
our guess is equivalent to the true model count. In this case, we print out the exact count and terminate the algorithm.\\
\textbf{\textit{Functions.}}
To motivate the definition of the function \texttt{ComputeCandK}, we view
an exhaust-up-to-$c$ query as analogous to measuring influence with a
bounded-length ``ruler.''
Suppose that we reduce the expected value of the model count by adding $k$ XOR constraints to $f$. 
Then, we can use the ``length-$(\log_2c)$ ruler''
to measure the influence starting at $k$ and this measurement corresponds to the result of an exhaust-up-to-$c$ query: the length-$(\log_2c)$ ruler has $c$ markings spaced logarithmically 
as illustrated in Figure \ref{fig:Ruler}.
Each iteration of the algorithm chooses a location ($k$) and length ($c$)
for the ruler, and gets a noisy reading on the influence as one mark on
the ruler.
Over time, we want to converge on the true influence value, but we also
wish to lengthen the rule so that the finer marks give more precise readings.
Based on this idea, we have the \texttt{ComputeCandK} function to choose the 
length of and starting point of the ruler from a prior distribution. 
Then, we run an exhaust-up-to-$c$ query and call \texttt{Update}
described in Section~\ref{sec:updating}
to update the distribution 
based on the result of the query. 

The pseudocode for algorithm \texttt{update} is 
described as Algorithm \ref{Update}.
A prior distribution $prior$ and a posterior distribution $post$ are represented as 
a set of sampled particles (influences). 
We sampled 500 particles for each \texttt{update} function call. 
Once we have the updated distribution, we can find out the interval of a given confidence level.\par

Since we observe that our running $\sigma$ represents how much we are close to the true answer, we use a rational function to satisfy 
the condition that $c$ increases as $\sigma$ decreases (i.e., we get more accurate result as $c$ increases).
\begin{algorithm}[!ht]
\caption{\texttt{Update}($prior$, $c$, $k$, $nSat$, $CL$)}\label{Update}
\begin{algorithmic}[1]
	\For{$t$ \texttt{from} $1$ \texttt{to} $nParticles$}
        \State $x_t \gets prior_t$
        \Statex{\Comment $x$ is a list of sampled particles (influences)}
        \If {$nSat < c$} 
        \Statex{\Comment Updating each weight of each particle}
            \State $w_t = Pr_{=nSat}(2^{x_t},k)$
        \Else
            \State $w_t = Pr_{\ge nSat}(2^{x_t},k)$
        \EndIf
    \EndFor
    \State $w \gets \texttt{normalize}()$
    \Comment Normalizing the weights
    \State $post\gets \texttt{sample}(x,w,nParticles)$
    \Statex{\Comment Resampling based on the weights}
    \State $UB,LB \gets \texttt{getBounds}(post, CL)$
    \State \Return $post,UB,LB$
\end{algorithmic}
\end{algorithm}

The $k$ value denotes where to put the ruler. We want to place the ruler where
the expected value of the prior distribution lies near the middle of the ruler hence our expected value is in the range of the ruler
with high probability.
Therefore, we subtract the half length of the ruler ($\frac{1}{2}\log_2c$) from the expected value $\mu$ and then use the
floor function to the value because $k$ has to be an nonnegative integer value. The expected value always lies in the right-half side of
the ruler by using the floor function. However, it is not essential which rounding function is used. Note that 
there might be a case where $k$ 
becomes negative. If this happens, we set $k=0$ and $c=\infty$, because our expected value is so small that we can run the solver
exhaustively to give the exact model count.
The formula for $c$ is motivated by the intuition that the spacing between two
marks near the middle of the ruler should be proportional to the standard deviation
of the the probability distribution, to ensure that a few different results of
the query are possible with relatively high probability; the spacing between the
two marks closest to $\frac{1}{2}\log_2c = \log_2{\sqrt{c}}$ will be about
$\log_2(\sqrt{c}+\frac{1}{2})-\log_2(\sqrt{c}-\frac{1}{2})$.
Setting this equal to $\sigma$, solving for $c$, and taking the ceiling gives
line 3 of Algorithm~\ref{ComputeCandK}.

\subsection{Probabilistic sound bounds}\label{sec:sound}
We have found that the binomial model performs well for choosing a series of
queries, and it yields an estimate of the remaining uncertainty in the tool's
results, but because the binomial models differs in an hard-to-quantify way
from the true probability distributions, the bounds derived from it do not
have any associated formal guarantee.
In this section we explain how to use our tool's same query results, together
with sound though less-precise bounds, to compute a probabilistically sound
lower and upper bound on the true influence.
As a trade-off, these bounds are usually not as tight as our tool's primary
results.

The idea is based on Lemma 2.10 and 2.13 from Klebanov \textit{et al.}'s work~\cite{QAPL2016}. 
They were inspired by \textsf{{ApproxMC}}~\cite{DBLP:conf/cp/ChakrabortyMV13} 
which is an $(\epsilon, \delta)$ counter such that the true model count is within the interval $[|F|/(1+\epsilon),|F|\cdot(1+\epsilon)]$ 
with a probability of at least $1-\delta$. 
They transformed the definition of the result interval to 
$[|F|\cdot(1-\epsilon),|F|\cdot(1+\epsilon)]$. 

\textbf{Lemma 1.} \textit{Let $n=|\Sigma|, \epsilon \in [0,1], k \in \mathbb{N}, k \leq \lfloor \log_2(|F|\cdot \epsilon^2 / (r \cdot \sqrt[3]{e}))\rfloor$ and $h \in $ H(n,k,r) a randomly chosen strongly r-universal hash function. It holds:}

\begin{flalign}
\nonumber Pr \left[(1-\epsilon) \cdot \frac{|F|}{2^k} \leq |F_h| \leq (1+\epsilon) \cdot \frac{|F|}{2^k} \right] \geq 1-e^{\lfloor -r/2 \rfloor} 
\end{flalign}

They called \textit{pivot} as what we call $c$ from an exhaust-up-to-$c$ query and $\textit{pivot}=\lceil \frac{2 \cdot r \cdot (1+\epsilon) \cdot \sqrt[3]{e}}{\epsilon^2} \rceil$. Since $0<\epsilon<1$ and $r=3$, $c$ (\textit{pivot}) should be greater than 17 to make the lemma true with a probability of at least 0.86 ($1-e^{\lfloor -3/2 \rfloor} \simeq 0.86$). 
In \textsf{{SearchMC}}'s iterations, given $c$ and $k$, we can compute $\epsilon$ value to estimate the bounds.
Therefore, when $c$ is greater than 17 from our tool's iteration, we can compute a lower and upper bound such that the true influence is within the bounds with a probability of at least 0.86.

\section{Experimental Results}\label{sec:result}
In this section, we present our experimental results. All our experiments were performed on a machine with an Intel Core i7 3.40Ghz CPU and 16GB memory.
Our main algorithm is implemented with a Perl script and \texttt{Update} function is implemented in a C program called by the main script.
Our algorithm can be applied to both SMT formulas and CNF formulas.
We have tested a variety of SAT solvers and SMT solvers, and our current implementation
specifically supports \textsf{Cryptominisat2}~\cite{Soos10enhancedgaussian,Soos:2009:ESS:1575471.1575502} %and \textsf{Cryptominisat4}~\cite{Soos10enhancedgaussian} 
for CNF formulas and \textsf{Z3}~\cite{DeMoura:2008:ZES:1792734.1792766} and \textsf{MathSAT5}~\cite{CimattiGSS2013:mathsat5} for SMT formulas.
We modified \textsf{Cryptominisat2} to count the number of solutions over specified variables.
For pure bit-vector SMT formulas, our tool also supports eagerly converting
the formula to CNF first and then using CNF mode.
(We implement the conversion using the first phase of the STP solver~\cite{GaneshD2007,STP}
with optimizations disabled and a patch to output the SMT-to-CNF variable
mapping.)
Performing CNF translation eagerly gives up the benefit of some (e.g.,
word-level) optimizations performed by SMT solvers, but it can sometimes
be profitable because it avoids repeating bit-blasting, and allows the
tool to use a specialized multiple-solutions mode of \textsf{Cryptominisat}.

% We first run our algorithm with random problems. 
% We uniformly generate 6400 sample problems over influence 0 to 64 similar to those used in table construction 
% and run our algorithm, by giving a 90\% confidence level and $thres=2$ , to check whether the influence lies between a
% lower bound and an upper bound computed by our algorithm. As a result, it shows the accuracy of 90.2\% which is close to the given 90\% confidence level.

\begin{figure*}[!ht]
\centering
\subfloat[Reported lower and upper bounds\label{fig:influence}]{%
  \includegraphics[trim={0.5cm 5cm 1cm 6cm},clip,width=.5\linewidth]{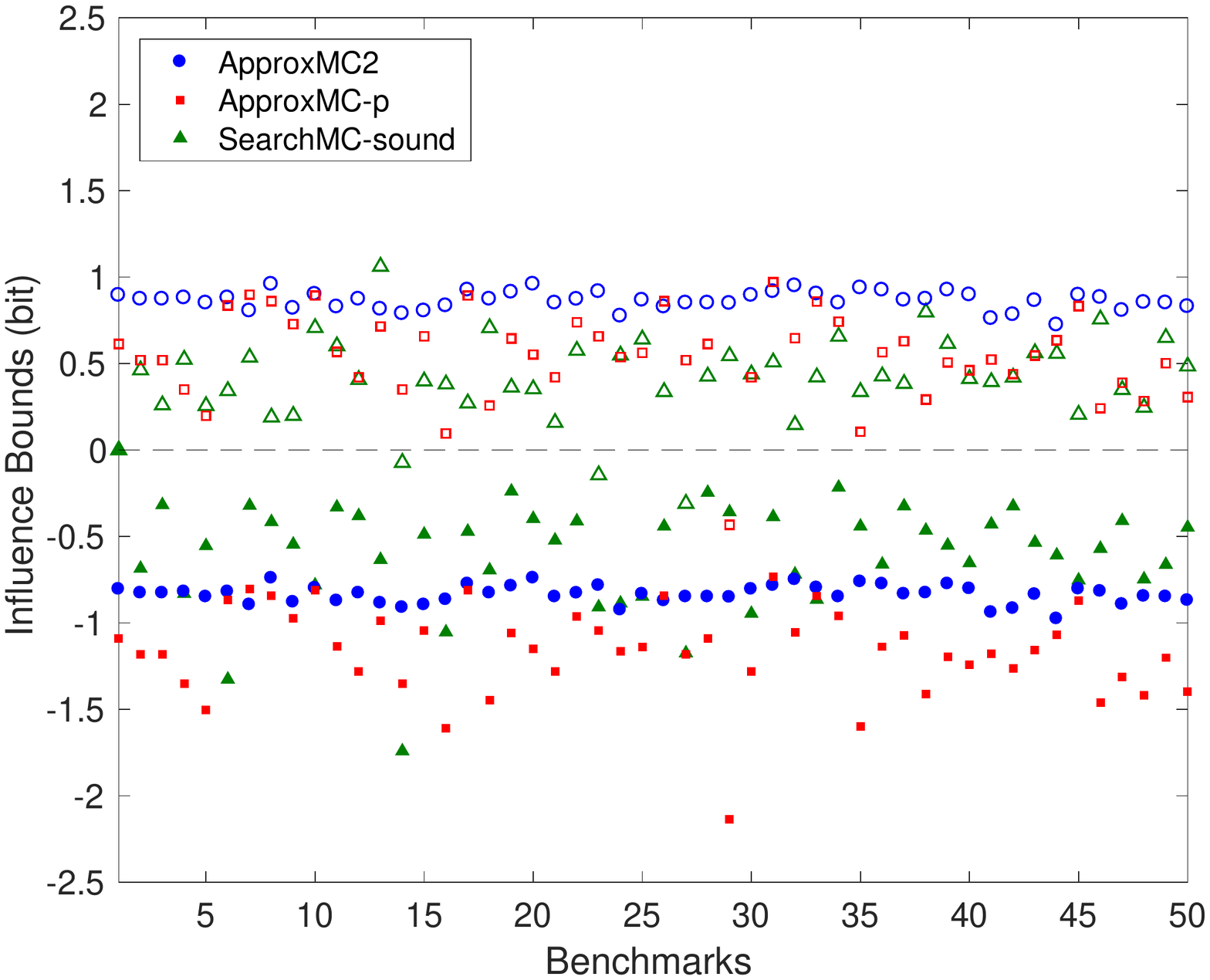}
}
~
\subfloat[Performance vs error trade-off\label{fig:Error}]{%
  \includegraphics[trim={0.5cm 5cm 1cm 6cm},clip,width=.5\linewidth]{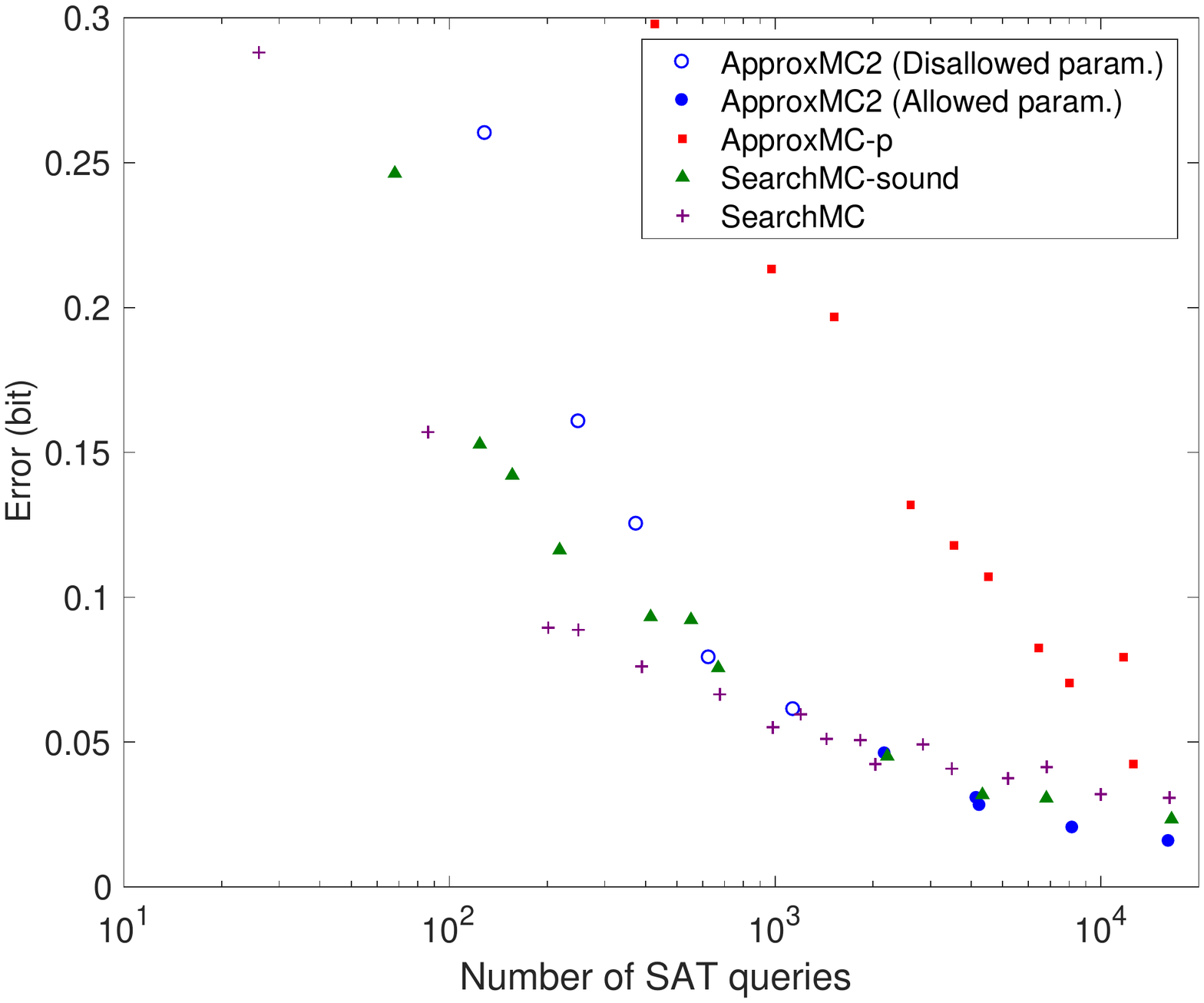}
}
\hfill
\subfloat[Time performance\label{fig:performance}]{%
  \includegraphics[trim={0.5cm 5cm 1cm 6cm},clip,width=.5\linewidth]{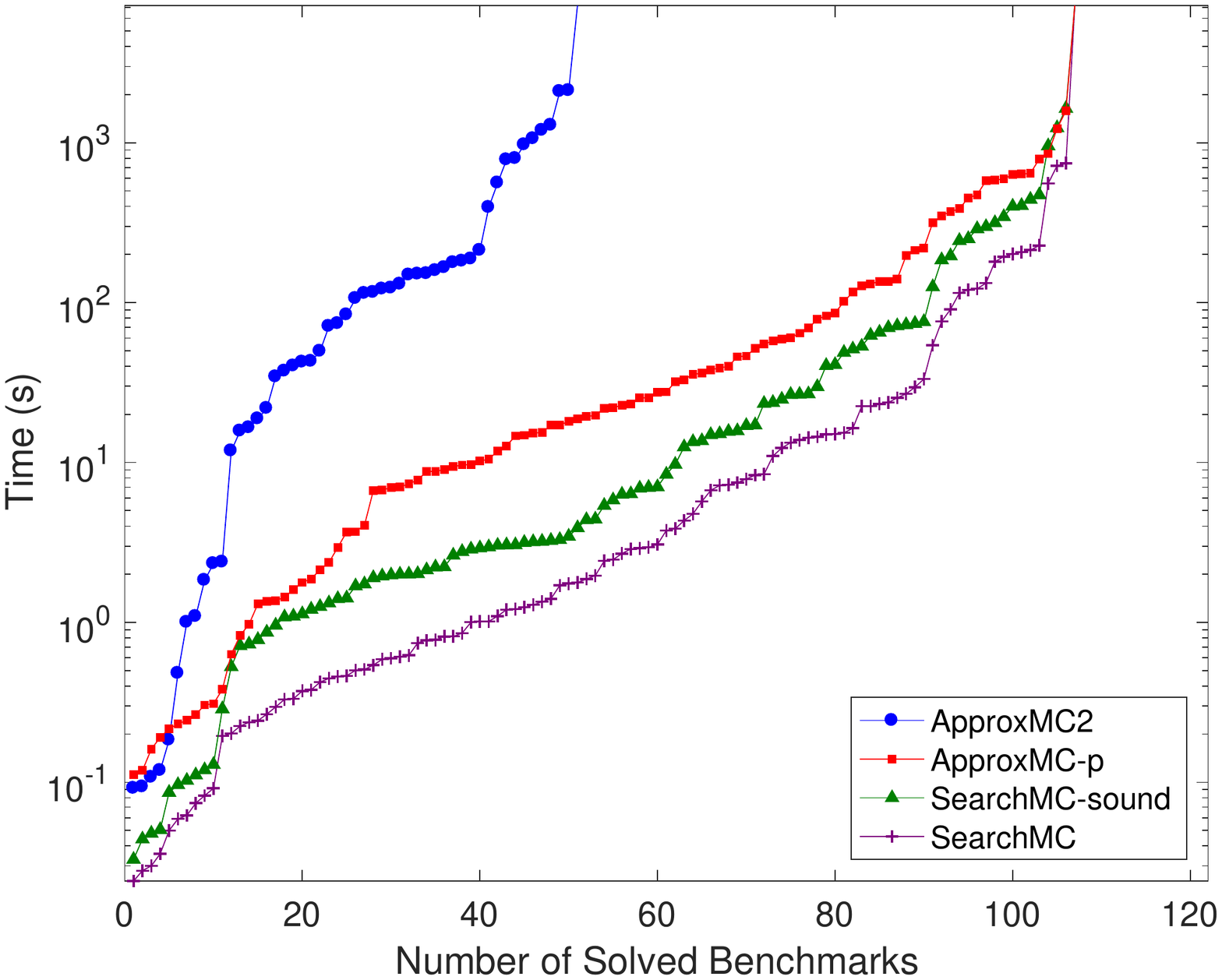}
}
~
\subfloat[Number of SAT queries\label{fig:SATquery}]{%
  \includegraphics[trim={0.5cm 5cm 1cm 6cm},clip,width=.5\linewidth]{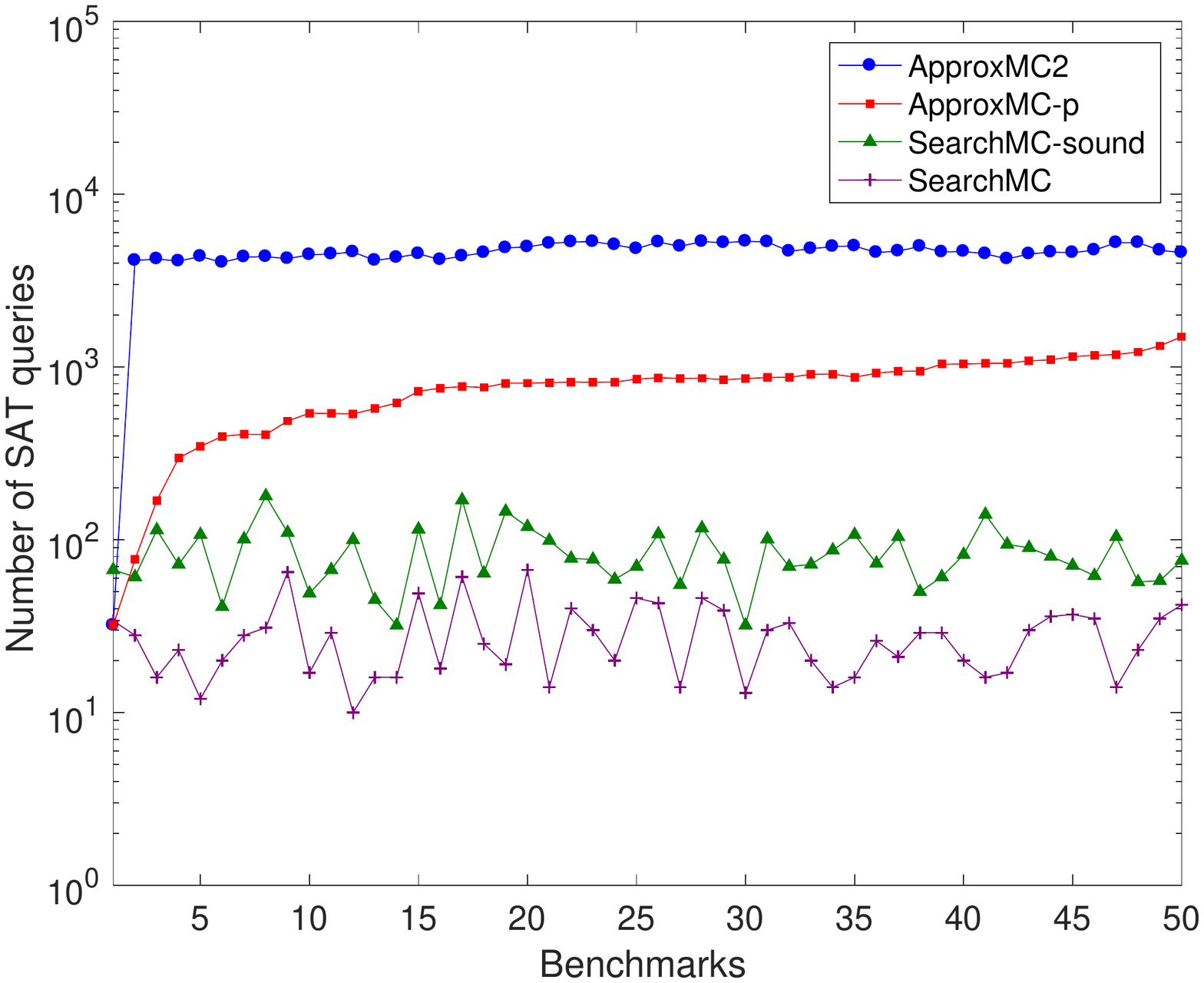}
}
\hfill

\caption{Comparison between \textsf{{SearchMC}} and \textsf{{ApproxMC}}s}\label{fig:comparison}
\end{figure*}

\subsection{Comparison with \textup{\textsf{ApproxMC2}} and \textup{\textsf{ApproxMC-p}}}\label{sec:cp}
We run our algorithm with a set of DQMR (Deterministic Quick Medical Reference) benchmarks~\cite{cachet} and ISCAS89 benchmarks~\cite{100747}
converted to CNF files by TG-Pro~\cite{IOPORT.06544917}
The DQMR benchmarks are medical diagnosis Bayesian networks and each problem is given by a two layer bipartite network graph for diseases and symptoms. There is an edge between the disease and the symptom if a disease
may result a symptom. In these benchmarks, the numbers of diseases and symptoms
are varied from 50 to 100 and each symptom is caused by four randomly chosen diseases. We want to compute the marginal probabilities for all the diseases given a set observations of symptoms when the size of observation set varies from
10\% to 30\% of all symptoms. The ISCAS89 benchmarks are digital sequential circuits that evaluate test pattern generation and TG-Pro can
convert them into CNF files. The benchmarks have from 7 to 24,843 variables and from 8 to 56,487 clauses.
%The non-decreasing benchmarks are that a set of sequences of numbers are made up of non-decreasing digits 
%(binary representation of a given bit size) for a given length of the sequences, 
%such that all the digits to the left of any digit is less than or equal to that digit. 
%For example, if we consider sequences that the length of the sequence is 4 and the bit size is 4, 
%one example sequence would be $\texttt{0x11ab}$ represented as hex digits. 
and compare the results of the benchmarks with \textsf{{ApproxMC2}}~\cite{CMV16} and \textsf{ApproxMC-p}~\cite{QAPL2016}. 
\textsf{{ApproxMC2}} and \textsf{ApproxMC-p} are state-of-the-art approximate \#SAT
solvers which we describe in more detail in Section \ref{sec:related}.
We used \textsf{Cryptominisat2} as the back-end solver with
all the tools for fair comparison.
For the parameters for the tools, we set $CL=0.86$ (confidence level),
$\alpha=0.5$ (confidence level adjustment factor) and 
$thres=1.7$ (desired interval length). 
As described above, \textsf{{SearchMC-sound}} gives correct bounds with a probability of at least 0.86.
Since the desired confidence level for \textsf{{ApproxMC2}} is $1-\delta$, it can 
achieve a 86\% confidence level by setting a parameter $\delta=0.14$ which corresponds to our parameter $CL=0.86$. 
The length of the interval for \textsf{{ApproxMC2}} is computed as 
$\log_2(|f|\times(1+\epsilon))-\log_2(|f|\times(1/(1+\epsilon)))=1.7$ hence we can obtain the interval length 1.7 
by setting a parameter $\epsilon=0.8$, corresponding to our parameter $thres=1.7$. 
Computing the interval for \textsf{ApproxMC-p} is a little different. The length of the interval for \textsf{{ApproxMC-p}} is $\log_2(|f|\times(1+\epsilon))-\log_2(|f|\times(1-\epsilon))=1.7$ hence we can obtain the interval length 1.7 
by setting a parameter $\epsilon=0.53$.
Note that \textsf{{SearchMC}} increases the $c$ value of 
an exhaust-up-to-$c$ query as it iterates while the corresponding \textsf{{ApproxMC2}} and \textsf{{ApproxMC-p}} parameters 
are fixed as a function of $\epsilon$ (72 and 46, respectively) in this experiment. Also, we set an initial prior to be a uniform distribution over 0 to 64 bits for \textsf{{SearchMC}}.
We tested 122 benchmarks (83 DQMRs and 39 ISCAS89s). 
\textsf{{SearchMC}} and \textsf{{ApproxMC-p}} were able to solve 106 benchmarks (83 DQMR and 23 ISCAS89) and \textsf{{ApproxMC2}}
was able to solve 50 benchmarks (35 DQMRs and 15 ISCAS89s) within 2 hours. The benchmarks that were solved completely by the other tools were also solved completely by \textsf{{SearchMC}}.
Figures \ref{fig:influence} and \ref{fig:SATquery} are based on the benchmarks that were solved by all tools. 

Figure \ref{fig:influence} compares the quality of lower bounds and upper bounds computed by \textsf{{SearchMC-sound}}, 
\textsf{{ApproxMC-p}} and \textsf{{ApproxMC2}}. The influence bounds are the computed bounds minus the true influence.
Filled markers and empty markers represent reported lower bounds and upper bounds, respectively.
Since \textsf{{ApproxMC2}} computes the bounds conservatively by using more queries, \textsf{{ApproxMC2}}'s intervals are more closely
centered on the true influence, and it out-performs the requested 86\%
confidence level.
By comparison, our tool and \textsf{{ApproxMC-p}} compute the bounds more aggressively, and in a few
cases the true result is just outside the reported interval (visible as empty markers below the dotted line), though this
still occurs somewhat less often than the 14\% implied by the confidence level.
\textsf{{SearchMC-sound}} tends to give tighter bounds than \textsf{{ApproxMC}}s 
since it stops when the interval length becomes less than $thres$,
while the interval lengths for \textsf{{ApproxMC}}s are fixed by a parameter $\epsilon$.

Figure \ref{fig:Error} shows another perspective on the trade-off
between performance and error.
We selected a single benchmark and varied the parameter settings of each
algorithm, measuring the absolute difference between the returned answer
and the known exact result.
We include results from running \textsf{{ApproxMC2}} with parameter
settings outside the range of its soundness proofs (shown as ``disallowed''
in the plot), since these settings are still empirically useful, and
\textsf{{SearchMC}} makes no such distinction.
From this perspective the tools are complementary depending on one's desired
performance-error trade-off.
The results from all the tools improve similarly with configurations that 
use more queries, but \textsf{{SearchMC}} performs best at getting more
precise results from a small number of queries.
About 100 satisfiability queries is a minimum for \textsf{{ApproxMC2}}
to give any results.
On the other hand our current \textsf{{SearchMC}} and \textsf{{SearchMC-sound}} implementation's results
improve as using more queries, the error varies over different runs compared to the stable 
error of \textsf{{ApproxMC2}}.

We also compare the running-time performance with \textsf{{ApproxMC}}s and show the running-time performance comparison on our 122 benchmarks in Figure \ref{fig:performance}. 
Since \textsf{{ApproxMC-p}} refined the formulas of \textsf{ApproxMC}, it used a smaller number of queries than \textsf{{ApproxMC2}}. \textsf{{SearchMC}} can solve all the benchmarks faster than \textsf{{ApproxMC}}s with 86\% confidence level.
\textsf{{SearchMC-sound}} performs faster than \textsf{{ApproxMC-p}} even \textsf{{SearchMC-sound}} computes its confidence interval similarly to \textsf{{ApproxMC-p}}.
%except only one benchmark. \textsf{{ApproxMC}}s queries the exhaust-up-to-$c$ and gives the exact influence immediately if the exact model count is less than $c$. One benchmark was applied to this case where the running time is very short.
The \textsf{{SearchMC}}'s and \textsf{{SearchMC-sound}}’s average running times are 41.79 and 84.89 seconds,
compared to an average of 127.8 for \textsf{{ApproxMC-p}}.
\textsf{{ApproxMC2}} requires an average of 281.25 seconds just for the subset
of benchmarks it can complete.

We also compare the number of SAT queries on the benchmarks for all the tools in Figure \ref{fig:SATquery}.
%One benchmark mentioned previously also used a very small number of queries.
The average number of SAT queries for \textsf{{SearchMC}}, \textsf{{SearchMC-sound}} \textsf{{ApproxMC-p}} and \textsf{{ApproxMC2}} is about 29.11, 79.75, 1257.19 and 4613.72 queries, respectively. 
Again this average for \textsf{{ApproxMC2}} is based on 50 benchmarks and others are based on 106 benchmarks.

\subsection{Floating Point / Differential Privacy Case Study}\label{sec:fp}
As an example of model counting with floating point constraints,
we measure the security of a mechanism for differential privacy
which can be undermined by unexpected floating-point behavior.
\begin{table*}[t]
\begin{small}
\begin{center}
\begin{tabular}{|c|r|c|r|c|r|}\hline
Problem        & \multicolumn{3}{|c|}{All noise} & \multicolumn{2}{|c|}{Intersection}\\
size           & Expected & SearchMC        & Time & SearchMC       & Time\\\hline
9e7, $2^6$     &  5.977 & 5.997   &   107s & 3.170          & 177s\\
10e7 $2^6$     &  5.977 & 5.997   &   158s & 3.459          & 290s\\
11e7, $2^6$    &  5.977 & 5.997   &   159s & 2.000          & 180s\\
12e7, $2^6$    &  5.977 & 5.997   &   145s & 2.000          & 264s\\
13e7, $2^7$    &  6.989 & [6.903, 7.505]  &  340s & 2.585 & 511s\\
14e7, $2^7$    &  6.989 & [6.709, 7.444]  &  338s & 2.000 & 454s\\
15e7, $2^8$    &  7.994 & [7.985, 8.813]  &  309s & 1.000 & 462s\\
16e7, $2^9$    &  8.997 & [8.678, 8.999]  &  1088s & 3.322 & 2668s\\
16e8, $2^{10}$ &  9.999 & [9.695, 10.292] &  1087s & 4.754 & 5831s \\
18e8, $2^{10}$ &  9.999 & [9.362, 10.288] &  739s & 1.000 & 1180s\\
19e8, $2^{11}$ & 10.999 & [10.700, 11.446] & 1940s & 3.585 & 6535s\\
\hline
\end{tabular}
\end{center}
\caption{Results and performance of model counting ($\log_2$ shown)
of naive Laplacian noise in IEEE floating point (Section~\ref{sec:fp})
\label{tab:laplace}}
\end{small}
\end{table*}
The Laplace mechanism achieves differential privacy~\cite{Dwork2006} by adding
exponentially-distributed noise to a statistic to obscure its exact value.
For instance, suppose we wish to release a statistic counting the number of
patients in a population with a rare disease, without releasing information
that confirms any single patient's status.
In the worst case, an adversary might know the disease status of all
patients other than the victim; for instance the attacker might know that the
true count is either 10 or 11.
If we add random noise from a Laplace distribution to the statistic before
releasing it, we can leave the adversary relatively unsure about whether
the true count was 10 or 11, while preserving the utility of an approximate
result.
A naive implementation of such a simple differentially private mechanism
using standard floating-point techniques can be insecure because of a
problem pointed out by Mironov~\cite{Mironov2012}.
For instance if we generate noise by dividing a random number
in $[1, 2^{31}]$ by $2^{31}$ and taking the logarithm, the relative
probability of particular floating point results will be quantized
compared to the ideal probability, and many values will not be
possible at all.
If a particular floating point number could have been generated as
$10 + \text{\em noise}$ but not as $11+\text{\em noise}$ in our
scenario, its release completely compromises the victim's privacy.

To measure this danger using model counting, we translated the standard
approach for generating Laplacian noise, including an implementation
of the natural logarithm, into SMT-LIB 2 floating point and bit-vector
constraints.
(We followed the {\tt log} function originally by SunSoft taken
from the {\tt musl} C library, which uses integer operations to reduce
the argument to $[\sqrt{2}/2, \sqrt{2})$, followed by a
polynomial approximation.)
A typical implementation might use double-precision floats with an 11-bit
exponent and 53-bit fraction, and 32 bits of randomness, which we
abbreviate ``$53e11, 2^{32}$'', but we
tried a range of increasing sizes.
We measured the total number of distinct values taken by
$10 + \text{\em noise}$ as well as the size of the intersection of
this set with the $11 + \text{\em noise}$ set.
%
%(To avoid a quantifier alternation, we first check that
%the intersection contains at least one element $x$ via a satisfiability
%query, then model count a formula of the form
%``if $10+n_1 = 11+n_2$ then $10+n_1$ else $x$''.)

The results and running time are shown in
Table~\ref{tab:laplace}.
We ran \textsf{{SearchMC}} with a confidence level of 80\%, a confidence level adjustment of 0.5 and a threshold of 1.0;
the SMT solver was \textsf{MathSAT 5.3.13}
with settings recommended for floating-point constraints by its authors.
We use one random bit to choose the sign of the noise, and the rest to
choose its magnitude.
The sign is irrelevant when the magnitude
is 0, so the expected influence for $n$ bits of randomness
is $\log_2(2^n - 1)$.
\textsf{{SearchMC}}'s 80\% confidence interval
included the correct result in all cases.
The size of the intersections is small enough that \textsf{{SearchMC}} usually
reports an exact result, and always much less than the total set of
noise values, confirming that using this algorithm and parameter
setting for privacy protection would be ill-advised.
The running time increases steeply as the problem size increases,
which matches the conventional wisdom that reasoning about
floating-point constraints is challenging.
But because floating-point SMT solving is a young area, there is
significant scope for future solvers to improve the technique's
performance.

\section{Related Work}\label{sec:related}
In this section, we summarize related work on model counting. We categorize model counting techniques into three areas: exact model counting, randomized approximate model counting and non-randomized approximate model counting. We also introduce some applications of model counting techniques for security and privacy purposes.

Exact model counters give the exact number of solutions for a formula but don't perform well as the size of a problem increases.
Approximate model counters have been proposed to resolve the scalability challenges. 
When a formula has a large number of solutions, 
it is often sufficient to provide rough estimates instead of the exact model counts, especially when this can be done much faster.
Randomized approximate model counting techniques are
likely to use random sampling and a SAT solver to produce a probabilistic result, which can often be argued to provide a lower bound and/or upper bound with high probability. In contrast, non-randomized approximate model counting techniques generally use approximations that are more efficient but do
not provide probabilistic bounds.

\subsection{Exact model counting}
Some of the earliest Boolean model counters used the DPLL algorithm~\cite{Davis:1962:MPT:368273.368557} for counting the exact number of solutions. 
Birnbaum \textit{et al.}~\cite{DBLP:journals/jair/BirnbaumL99} formalized this idea and introduced an algorithm
for counting models of propositional formulas. Based on this idea, \textsf{{Relsat}}~\cite{Bayardo:1997:UCL:1867406.1867438},
\textsf{{Cachet}}~\cite{DBLP:conf/sat/SangBBKP04} \textsf{{sharpSAT}}~\cite{Wei:2005:NAM:2129929.2129954}
and \textsf{{DSHARP}}~\cite{Muise_dsharp}
showed improvements by using several optimizations. 
\textsf{{Relsat}} uses component analysis in which the 
model count of a formula is the product of the model count of each sub-formula (component).
\textsf{{Cachet}} shows optimizations by combining component caching and clause learning.
\textsf{{sharpSAT}} introduces an improved caching technique to reduce the space requirement compared to \textsf{{Cachet}}.
\textsf{{DSHARP}}~\cite{Muise_dsharp} is a CNF $\rightarrow$ d-DNNF compiler for efficient reasoning and uses \textsf{{sharpSAT}} as a back-end.
The major contribution of \textsf{{countAntom}}~\cite{Burchard2015} is techniques for parallelization, but it provides state-of-the-art
performance even in single-threaded mode.

Phan \textit{et al.}~\cite{Phan:2012:SQI:2382756.2382791} encode a full binary search for feasible outputs in a bounded model checker. 
This approach is precise, but requires more than one call to the underlying solver for each feasible output. 
Specifically, it recursively calls the solver by adding a bit constraint for finding a single satisfying assignments as DPLL-based search.
This is most similar to the exhaust-up-to-$c$ approach that requires $n+1$
queries to find $n$ feasible outputs.
This search tree approach is useful when the program verification system does not expose the underlying logical representation or when the used solver 
cannot generate models.
Klebanov \textit{et al.}.~\cite{KlebanovMM2013} perform exact model counting
for quantitative information-flow measurement, with an approach that converts
C code to a CNF formula with bounded model checking and then uses exact
\#SAT solving.
They explore both exhaustive enumeration and the existing \textsf{DSHARP} 
and \textsf{sharpSAT} tools, but only counting distinct values of the
output variables.
Val \textit{et al.}~\cite{ValEBAH2016} integrate a symbolic execution tool
more closely with a SAT solver by using techniques from SAT solving to prune
the symbolic execution search space, and then perform exact model counting
restricted to an output variable.\\
However, precisely counting solutions is a \#P-complete problem and these exact model counters typically 
only work well with small-sized problems or ones with only simple constraints.
For many practical problems, it is infeasible
to count the exact number of solutions in a
reasonable amount of time.
% Klebanov et al. also show an image anonymization example, but measure
% information flow not of pixel values but of pixel positions: their results
% show that if an image is transformed by a discretized version of a
% continuous swirl, the location of an output pixel reveals most of the
% information about the corresponding input pixel location.

\subsection{Randomized approximate model counting}
Randomized approximate model counting techniques perform well on many kinds of a formula for which finding single solutions is efficient.
Also, there can sometimes be a smooth trade-off chosen between computational 
effort and the precision of results.
However, this solving is still relatively expensive, so research to get the
best precision for a given cost is still important.
Wei and Selman~\cite{Wei:2005:NAM:2129929.2129954} introduced \textsf{{ApproxCount}} which uses near-uniform sampling to estimate the true model count but it can significantly over-estimate or underestimate if the sampling is biased. 
\textsf{{SampleCount}}~\cite{Gomes07fromsampling} improves this sampling idea and gives a lower bound
with high probability by using a heuristic sampler. 
\textsf{{MiniCount}}~\cite{Kroc_leveragingbelief} is based on a framework to compute an upper bound under statistical assumptions which is that 
counting the number $d$ of branching decisions (except unit propagations and failed branches) can be used to estimate the total number of solutions by setting a variable to true or false randomly. 
Specifically, they showed that 
the expected value of $d$ is not
lower than $\log_2$ of the true model count. By estimating the expected value, they
can obtain the upper bound of the true model count.
Also, they observed that $d$ often has a distribution that is close to a normal distribution. Thus, the expected value can be easily computed under this assumption rather than computing low and high values of $d$.
This only guarantees the upper bound and they used
a different method to compute a lower bound.

\textsf{{MBound}}~\cite{DBLP:conf/aaai/GomesSS06} is an approximate model counting tool that gives probabilistic bounds on the model
counts by adding randomly-chosen parity constraints as XOR streamlining. 
Chakraborty \textit{et al.}~\cite{DBLP:conf/cp/ChakrabortyMV13} 
introduced \textsf{{ApproxMC}}, an approximate model counter for CNF formulas, which automated the choice of XOR streamlining parameters.
The \textsf{{ApproxMC}} algorithm, in our terminology, starts by fixing
$c$ and a total number of iterations based on the desired precision and
confidence of the results.
In each iteration \textsf{{ApproxMC}} searches for an appropriate $k$ value,
adds $k$ XOR random constraints, and then performs an exhaust-up-to-$c$ 
query on the streamlined formula and multiplies the result by $2^k$.
It stores all the individual estimates as a multiset and computes its
final estimate of the model count as the median of the values.
The original \textsf{{ApproxMC}} sequentially increases $k$ in each iteration 
until it finds an appropriate $k$ value. 
An improved algorithm \textsf{{ApproxMC2}}~\cite{CMV16} uses galloping
binary search and saves a starting $k$ value between iterations to make the
selection of $k$ more efficient.
Other recent systems that build on \textsf{{ApproxMC}} include
\textsf{{SMTApproxMC}}~\cite{CMMV15} and
\textsf{{ApproxMC-p}}~\cite{QAPL2016}.
\textsf{{SMTApproxMC}} proposes word-level constraints based on modular 
arithmetic instead of bit-level XOR constraints, however these will not
likely provide comparable performance until SMT solvers implement
modular-arithmetic Gaussian elimination.
\textsf{{ApproxMC-p}} implements projection (counting over only a subset of 
variables), and also gives more efficient formulas for parameter selection.

\textsf{{ApproxMC2}}, developed concurrently, is the system most similar
to \textsf{{SearchMC}}: its binary search for $k$ plays a similar role
to our converging $\mu$ value.
However \textsf{{SearchMC}} also updates the $c$ parameter over the course
of the search, leading to fewer total queries.
\textsf{{ApproxMC}}, \textsf{{ApproxMC2}}, and related systems choose the
parameters of the search at the outset,  and make each iteration either
fully independent (\textsf{{ApproxMC}}) or dependent in a very simple way
(\textsf{{ApproxMC2}}) on previous ones.
These choices make it easier to prove the tool's probabilistic
results are sound, but they require a conservative choice of parameters.
By comparison \textsf{{SearchMC}}'s approach of maintaining a probabilistic
estimate explicitly at runtime means that its iterations are not at
all independent: instead our approach is to extract the maximum guidance
for future iterations from previous ones, to allow the search to
converge more aggressively.

The runtime performance of \textsf{{SearchMC}}, like that of \textsf{ApproxMC}(2),
is highly dependent on the performance of SAT solvers on CNF-XOR formulas.
Some roots of the difficulty of this problem have been investigated by Dudek
et al.~\cite{DMV16,DMV17}.

\subsection{Non-randomized approximate model counting}
Non-randomized approximate model counting using techniques similar to
static program analysis is generally faster than 
randomized approximate model counting techniques, and such systems can
give good approximations for some problem classes.
However, they cannot provide a precision guarantee for arbitrary problems,
and it is not possible to give more effort to have more refined results.
 
Castro \textit{et al.}~\cite{Castro:2008:BBR:1346281.1346322}
compute an upper bound on the number of bits about an input that are revealed
by an error report. 
They measure the entropy loss of an
error report by computing the number of bits revealed by subsets of path conditions
first and then combining these partial results to get the final result.
Meng and Smith~\cite{Meng:2011:CBI:2166956.2166957} use two-bit-pattern SMT entailment queries to calculate a
propositional overapproximation and count its instances with a model counter
from the computer algebra system Mathematica.
Luu \textit{et al.}~\cite{Luu:2014:MCC:2594291.2594331} propose a
model counting technique over an expressive string constraint language.
Their tool computes
the bounds on the cardinality of the valid string set
and uses generating functions for reasoning 
about the cardinality of string sets.

\subsection{Applications: Security and Privacy}
Various applications of model counting have been proposed for security and privacy 
purposes.
For instance Castro \textit{et al.}~\cite{Castro:2008:BBR:1346281.1346322}
use model counting and symbolic execution approaches to measure leaking private information from bug reports. They compute an upper bound on the amount of private
information leaked by a bug report and allow users to decide on whether
to submit the report or not.
Newsome \textit{et al.}~\cite{Newsome:2009:MCC:1554339.1554349} show how an untrusted input affect a program and 
introduce a family of technique for measuring influence which can be applicable to x86 binaries. 
They applied the XOR streamlining and the exhaust-up-to-$c$ techniques separately, but did not combine the two techniques
or describe an algorithm for choosing XOR streamlining parameters.

\section{Future Work}
\label{sec:future}

%With the particle filter approach, it is not necessary to have a fixed upper
%bound such as $2^{64}$ on the solution count: this could be a parameter,
%or the initial prior distribution could even be unbounded.
%
% However some numeric implementation issues such as the computation of
% large binomial coefficients will need special attention when solution
% counts become large relative to machine integers or floating-point
% precision.

%Proving a (probabilistic) soundness result for a variation of our technique
%seems challenging: for instance the adaptive nature of the algorithm
%means that the iterations can not be reasoned about as statistically
%independent, and a full statement would need to take into account the
%size of the particle filter.
%
%The most fundamental obstacle to soundness is the limited independence
%of our XOR constraints and the resulting imprecision in the probability
%model: adjusting the confidence is practically useful but it would be
%more satisfying to formally bound the error this imprecision introduces.

%It would be better to talk about improvements to the sound bounds
%technique instead here.

Closing the gap between the performance of \textsf{SearchMC} and
\textsf{SearchMC-sound} is one natural direction for future research.
On one hand, we would like to explore techniques for asserting sound
probabilistic bounds which can take advantage of the results of all of
\textsf{SearchMC}'s queries.
At the same time, we would like to find a model of the number of solutions
remaining after XOR streamlining that is more accurate than our current
binomial model, which should improve the performance of \textsf{SearchMC}.
Another future direction made possible by the particle filter implementation
is to explore different prior distributions, including unbounded ones.
For instance, using a negative exponential distribution over influence as a
prior would avoid the any need to estimate a maximum influence in advance,
while still starting the search process with low-$k$ queries which are typically
faster to solve.

\section{Conclusion}
\label{sec:conclusion}
In sum, we have presented a new model counting approach \textsf{{SearchMC}} using XOR 
streamlining for SMT formulas with bit-vectors and other theories.
We demonstrate our algorithm that adaptively maintains a probabilistic model count estimate 
based on the results of queries. 
Our tool computes a lower bound and an upper bound with a requested confidence level,
and yields results more quickly than previous systems.

\section*{Acknowledgement}
This research is supported by the National Science Foundation 
under grant no. 1526319. Any opinions, findings, and conclusions
or recommendations expressed in this material are those of the
authors and do not necessarily
reflect the views of the National Science Foundation.

\bibliographystyle{abbrv}
\bibliography{main}

\end{document}